# Optical Waveguide-Pair Design for CMOS-Compatible Hybrid III-V-on-Silicon Quantum Dot Lasers

**Peter Raymond Smith [1], Konstantinos Papatryfonos [1,2] and David R. Selviah [1,*]**

[1] Department of Electronic and Electrical Engineering, University College London, UCL, London WC1E 7JE,
[2] Institute of Electronics, Microelectronics and Nanotechnology (IEMN), UMR CNRS 8520, University of Lille, Avenue Poincare, 59650 Villeneuve of Ascq, France;
[*] Correspondence: d.selviah@ucl.ac.uk ; konstantinos.papatryfonos@iemn.fr

**Abstract**

The development of compact, energy-efficient integrated lasers operating at 1.3 μm remains a critical focus in silicon photonics, essential for advancing data communications and optical interconnect technologies. This paper presents a numerical study of distributed Bragg reflector (DBR) hybrid III-V-on-silicon lasers, analyzing design trade-offs and optimization strategies based on supermode theory. The III-V section of the design incorporates InAs/(Al)GaAs quantum dots (QDs), which offer improved temperature insensitivity at the cost of more complex III-V/Si optical coupling, due to the high refractive index of (Al)GaAs. Consequently, many current laser designs rely on silicon waveguides with a thickness exceeding 220 nm, which helps coupling but limits their compatibility with standard CMOS technologies. To address this challenge, we perform detailed simulations focusing on 220-nm-thick silicon waveguides. We first examine how the mode profiles jointly depend on the silicon waveguide dimensions and the geometry and composition of the III-V stack. Based on this analysis, we propose a novel epitaxial design that enables effective III-V/Si coupling, with the optical mode strongly confined within the III-V waveguide in the gain section and efficiently transferred to the silicon waveguide in the passive sections. Moreover, the final design is shown to be robust to fabrication-induced deviations from nominal parameters.

## 1. Introduction

The exponential growth of hyperscale data centers (DCs) designed for cloud applications, artificial intelligence (AI/ML) training and inference, social media, and big data analytics is driving advancements in the datacom industry, requiring high-speed servers, switches, and optical interconnects [1, 2]. The increasing demand for data throughput in modern information processing requires unprecedented bandwidth and low power consumption. Silicon Photonics (SiP) has emerged as a promising platform to address these challenges [3]. SiP-based interconnects are rapidly becoming the leading technology in the datacom environment due to their inherent benefits, including low-cost manufacturing, low power consumption, low crosstalk, and most importantly, the potential for integration with mature complementary metal-oxide-semiconductor (CMOS) technology. The recent adoption of such photonic technologies in DCs has resulted in significant performance enhancements [4]. However, the rapid increase in data traffic is outpacing current technologies, requiring innovative approaches at both the component [6, 7] and system levels [2, 5].

The SiP platform leverages the extensive silicon infrastructure of the CMOS industry by integrating silicon with other materials to enable combined functionalities, with particular emphasis on integrated III-V light sources [6, 2] and modulators [2, 7]. III-V-on-Si laser integration specifically aims to overcome silicon's limitations in light emission due to its indirect bandgap. A primary goal is therefore the implementation of efficient and reliable lasers [6, 8-10], which are crucial for many datacom applications. More specifically, high-performance optical interconnects for hyperscale DCs are a key application of SiP, and, thus, achieving integrated lasers with low threshold currents, low temperature sensitivity, and appreciable output powers even at elevated temperatures is of paramount importance [6]. Such advancements would enable the replacement of external light sources, paving the way for fully integrated architectures.

Currently, the main platforms striving for optimal III-V-on-Si laser integration are direct growth [5, 6, 9, 10] and bonding [8, 11, 12], which includes both device bonding after fabrication and wafer bonding prior to III-V processing [8]. In this paper, we focus on the latter approach. One promising pathway is the development of hybrid III-V/Si lasers using InAs/GaAs quantum dots (QDs). In a hybrid laser, light is generated and amplified in the III-V stack and then it couples down, by evanescent coupling, into a parallel silicon waveguide from which it can be sent to other SiP components. The coupling is controlled by changing the geometry of the III-V and Si waveguides in different segments of the laser.

Various hybrid laser devices have shown promising performance [2, 14-27], including Fabry-Pérot lasers [2, 6, 17], ring lasers [14], DFB [15] and DBR [16, 18] lasers. In particular, the recent demonstration of a DBR hybrid laser with a transfer printed InAs/GaAs quantum dot amplifier, highlights the potential for developing hybrid integrated laser devices that leverage the superior properties of QDs as the gain medium [28]. However, current CMOS technologies use 220 nm thick silicon waveguides [2, 10], limiting the usefulness of many of the designs mentioned above as they contain thicker silicon waveguides [14-28] that are not compatible with most CMOS processes. To maximize compatibility with CMOS technology for large scale manufacturing, a 220 nm silicon waveguide would be highly preferable [2], and it is becoming a high priority in recently developed Silicon Photonics (SiP) platforms [2, 10].

In this paper, the silicon device layer is fixed at 220 nm to align with CMOS-compatible SiP platforms. While thicker Si waveguides (>220 nm) can ease III–V/Si coupling and reduce associated losses, they are non-standard in foundry process design kits (PDKs) and, thus, fall outside the scope of our targeted integration path. Our optimization, therefore, focuses on the III–V epitaxial stack and waveguide cross-section under the 220-nm thickness constraint, which reflects the intended low cost, mass manufacturing route. The main difficulty in using such thinner Si waveguide in hybrid lasers, comes from the relative reduction of the Si waveguide's effective refractive index, which strongly depends on its thickness at values around 220 nm. This in turn imposes further constraints on the design of the III-V stack. In this study, we perform a numerical investigation aimed at developing a design method that accounts for the trade-offs involved in achieving optimal hybrid laser performance.

We focus on a DBR design and use the layer structure shown in Table 1 as the starting point. This structure utilizes InAs quantum dots (QDs) emitting at 1.3 µm as the gain medium. In recent years, epitaxial InAs QDs grown on both GaAs and Si substrates for 1.3 µm emission [6, 13, 29, 30, 31], as well as InAs quantum dashes (Qdashes) [32-34] on InP for 1.55 µm and above, have shown notable lasing performances [6, 29, 31, 36, 36]. Of relevance to our project is the high degree of temperature insensitivity shown by these lasers [6, 37, 38], attributed to their lower density of states (DOS), as recently shown experimentally using scanning tunneling spectroscopy [6, 32] confirming pioneering theoretical predictions [39, 40]. This feature is especially helpful for hybrid lasers in data center applications, where elevated intrinsic device temperatures are common.

Additionally, due to the QD quasi-zero-dimensional nature and delta-function-like DOS, QD lasers have low threshold current densities, even when size inhomogeneity arises from their self-assembled epitaxial growth. Moreover, QDs are less sensitive to defects than QWells allowing them to be grown on a wider variety of substrates [6, 29].

**Table 1.** Initial epitaxial structure

|  | Material | Thickness | n | Repeats |
|---|---|---|---|---|
| p-type | GaAs p-contact 1e18 | 50 nm | 3.412 | |
| Cladding | $Al_{0.4}Ga_{0.6}As$ | 1400 nm | 3.212 | 1 |
| | GaAs waveguide | 70 nm | 3.412 | |
| | $In_{0.18}Ga_{0.82}As$ capping | 2 nm | 3.434 | |
| Active Region | InAs | 2.7 ML | 3.532 | 5 |
| | $In_{0.18}Ga_{0.82}As$ capping | 6 nm | 3.434 | |
| | GaAs barrier | 37.5 nm | 3.412 | |
| | GaAs waveguide | 32.5 nm | 3.412 | |
| n-type | GaAs n-contact 5e18 | 50 nm | 3.412 | |
| Bonding Layer | $Al_2O_3$ | 30 nm | 1.75 | |
| | $SiO_2$ | 100 nm | 1.45 | 1 |
| | **Si Waveguide** | 220 nm | 3.5 | |
| SOI | $SiO_2$ | 500 nm | 1.45 | |
| | **Si Substrate** | 300 μm | 3.5 | |

## 2. Materials and Methods

<u>The supermode theory</u>: As shown by Yariv and Sun [41-44], a system of parallel evanescently coupled optical waveguides can be described by supermode theory. Supermode theory assumes that each mode that exists in the coupled system is a linear combination of the uncoupled modes of the waveguides. For each pair of modes of the individual waveguides, there exist two modes of the coupled system. These are known as the odd and even modes (subscripts o and e); the electric field is given by Equation (1) [41]:

$$E_e(x,y,z) = \left[ u_a + \frac{i\kappa_c^*}{\delta - S} u_b \right] e^{-i(\bar{\beta}+S)z},$$

$$E_o(x,y,z) = \left[ u_a + \frac{i\kappa_c^*}{\delta + S} u_b \right] e^{-i(\bar{\beta}-S)z}, \qquad (1)$$

where

$$\bar{\beta} = \frac{\beta_a' + \beta_b'}{2} \quad , \quad \delta = \frac{\beta_b' - \beta_a'}{2} \quad , \quad S = \sqrt{\delta^2 + \kappa_c^2} \quad ,$$

$u_a$, $u_b$ are the eigenmodes of the individual uncoupled waveguides, and $\beta_a'$, $\beta_b'$ are the propagation constants of the coupled waveguides.

Two key parameters in these equations are, δ, and, $\kappa_c$. The coupling coefficient, $\kappa_c$, is given in Equation (2) by the overlap integral of the two constituent modes of the supermode and the index perturbations [40]. It, thus, gives a measure of the coupling strength between the waveguides. By tuning the parameter, δ, we can determine in which of the two waveguides the supermode's power is more concentrated. To tune, δ, we need to change the effective refractive indices of the waveguides [45], which we achieve by modifying either their geometry—height and width, or the materials composing and surrounding the waveguides.

If the width of one or both waveguides is changed adiabatically e.g., they are slowly widened along the x-direction of Fig. 1a, it is possible to transfer the power of a mode from one waveguide to the other without significant scattering into other modes. In this way, the waveguide widths in the amplification region of the laser can be designed to confine most of the light in the III-V gain stack, maximizing the gain. Then, at the output of the laser, the waveguide widths can be changed to transfer it into the silicon waveguide, thus, giving maximum output power for a given gain. Supermode lasers of this type have been shown experimentally [46] to have superior characteristics to those in which the waveguide widths are kept constant. Since these devices can have significantly lower threshold currents and higher slope efficiencies to those with constant width waveguides, such devices can be smaller and have lower power consumption [42]. We therefore concentrate on designs of this type.

Simulation and optimization of realistic device structures using Supermode Theory: To apply the supermode theory to our system, a few approximations are necessary. First, we utilize Marcatili's method, which recently has been extended to enhance its accuracy for high-index-contrast (HIC) waveguides [47]. This approach allows us to find the eigenmodes of our designed waveguides, taking into account their width and the surrounding materials. Second, we approximate the multilayer active region as a single homogeneous waveguide, using an effective refractive index equal to the weighted average of the individual layer indices. In earlier modeling of similar structures, we found that this approximation significantly simplifies calculations with negligible impact on accuracy [32]. We note that while the active region is formed of many layers, it is mostly composed of $Al_xGa_{(1-x)}As$. Since the electric field must be continuous and differentiable everywhere, the thin quantum dot layers which have a slightly higher refractive index will not significantly change the shape of the mode, justifying this approximation.

Accurate knowledge of the refractive index for each $Al_xGa_{(1-x)}As$ layer is required. Gehritz et al. [48] developed a model for determining the refractive index of $Al_xGa_{(1-x)}As$ across different compositions. Given the significant role of GaAs and $Al_xGa_{(1-x)}As$ in a wide range of applications [2, 6, 49-54], recent experimental work with improved accuracy in the composition range x < 0.45 has been reported, aiming to improve the precision of numerical simulations for these devices [55]. We note that our design is particularly sensitive to refractive index inaccuracies. Specifically, our study (Section 3.3 and Fig. 5) shows that a 1% deviation in a refractive index value can alter the mode confinement factor by up to 30% in the cladding (Fig. 5f), and by about 20% in the waveguide and spacer layers (Fig. 5d), thus, underscoring the need for highly precise refractive index measurements. In our study, we use the accurate experimental data from Ref. [55] for compositions up to x=0.45, and for higher x > 0.45, we rely on the Gehritz et al. model from Ref. [48].

With these accurate refractive index values at hand, we aim to investigate the precise impact of varying the $Al_xGa_{(1-x)}As$ composition on mode profiles and confinement factors within the III-V and Si waveguides. Specifically, we will perform a series of simulations gradually increasing the value of the composition parameter, $x$, both in the active region starting from x=0 as well as in the cladding layers starting from x=0.4 (see initial design in Table 1), until we reach an optimal design in terms of mode profiles. We note that this design will be further refined to account for trade-offs related to material and fabrication limitations, as discussed in Sections IV and V. Nevertheless, these initial simulations are essential for establishing the acceptable ranges of the composition parameter, *x*, in each $Al_xGa_{(1-x)}As$ layer.

To achieve this, we develop a numerical approach that combines a simplified 2D model—used to guide a rapid parameter search—with a more sophisticated 3D model for detailed investigation within the parameter ranges identified by the 2D model. Such numerical studies of key parameters—enabled by combining a computationally efficient model, based on well-justified simplifications, with a more comprehensive model for detailed investigation within a narrowed parameter space—serve as a powerful tool for understanding device behavior and revealing underlying physical trends in diverse material platforms and applications [53, 54, 56-60, 61]. In this spirit, and with the above considerations in mind, we begin by developing the simplified numerical model of the system in MATLAB.

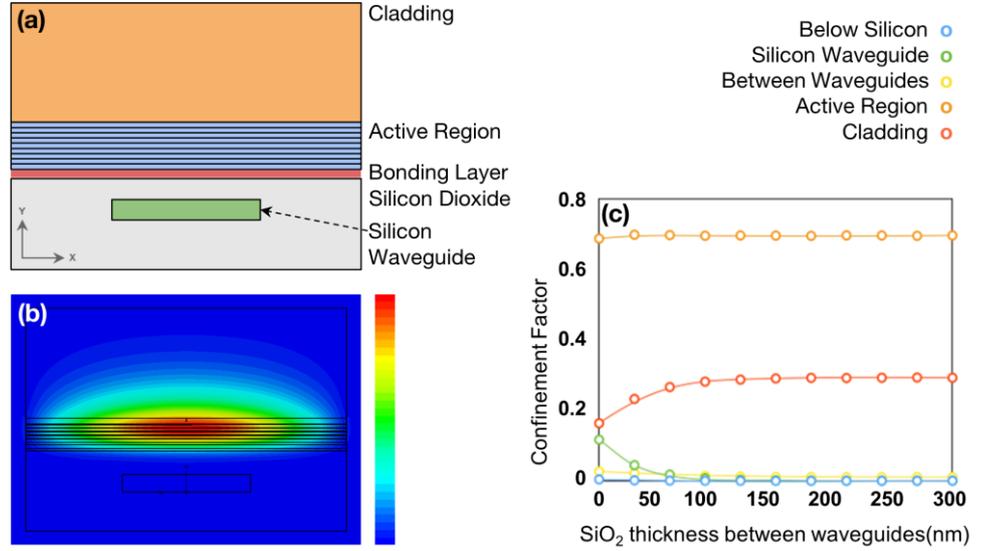

**Figure 1.** (a) CST model of evanescently couped waveguide structure. In this model, the full width of the structure is 4 μm. (b) Fundamental eigenmode of initial design calculated by this model with a 1.5 μm wide silicon waveguide and an overall width of 4 μm. (c) Dependence of confinement factors for each region in the structure on thickness of silicon dioxide bonding layer between the III-V stack and the silicon waveguide, calculated using supermode theory.

The first step involves applying the extended Marcatili's method and calculating the eigenmodes of the individual waveguides—the III-V and silicon waveguides. These modes are then used in the supermode equations (Eq. (1)) to find the 1D mode distribution between the two waveguides in each region of interest. Specifically, by integrating the field intensity vertically (along the y-axis as seen in Fig. 1a) over each region of interest, we obtain a measure of the proportion of the supermode confined to each waveguide. We, henceforth, refer to this quantity as the confinement factor. We then use this model to investigate how each design parameter (waveguide widths, material composition in each layer) affects the confinement factor between the III-V and Si waveguides in each region of the structure (Eq 2).

The modal field distributions and the phase velocities of the supermodes are determined by the coupling coefficient, $\kappa_c$, (computed from the field-overlap integral) and the phase mismatch parameter, $\delta$, between the modes of the uncoupled waveguides, as given in Eq. (2) for a 2D geometry [43, 44]:

$$\kappa_c = \sqrt{\kappa_{ab}\kappa_{ba}}, \text{ and}$$

$$\delta = \frac{(\beta_b + M_b) - (\beta_a + M_a)}{2}, \qquad (2)$$

where

$$\kappa_{ab.ba} = \frac{\omega \varepsilon_0}{4} \int_{-\infty}^{\infty} [n_c(x)^2 - n_{b,a}^2(x)] u_a(x) u_b(x) dx,$$

$\beta_a$, and $\beta_b$ are the propagation constants of the uncoupled waveguide modes, $n_a(x)$, $n_b(x)$, and $n_c(x)$ are the refractive index profiles of the uncoupled waveguides a and b and the coupled waveguide system c, respectively. $M_a$ and $M_b$ are small corrections to the propagation constants $\beta_a$, and $\beta_b$, respectively, due to the presence of the other waveguide [43, 44].

To complement and refine these results, we then develop a full 3D model using the software CST [62], which employs finite-difference time-domain (FDTD) algorithms to simulate optical behavior very accurately. The cross-section of the 3D CST model is illustrated schematically in Fig. 1a. The main results are presented in Sections IV and VI and illustrated in Figs. 2 and 4, while the silicon waveguide structure is described in more detail in Fig. 3 and Section V.

## 3. Results

### 3.1. Implementation of Supermode Theory in Hybrid Laser Design

The most practical and effective way to change the value of, δ, (see Eq. (1)) and, hence, the mode distribution, is to vary the width of the Si waveguide. Ideally, all other parameters of the system should be chosen such that the range of, δ, values necessary to achieve the desired mode distributions in each section, set by, $\kappa_c$, can be accessed by only varying the Si waveguide width. We start by simulating the structure shown in Table 1.

This structure uses a quantum dot-in-a-well (Dwell) gain stack, which was recently employed to demonstrate high-performance and exceptionally temperature-tolerant Fabry-Pérot single transverse mode lasers through direct epitaxial growth on silicon [6]. In this case, we aim to design a similar structure to that, but with wafer-bonding replacing direct growth on silicon, which would be preferable for certain applications. The 300 nm thickness of the $SiO_2$ bonding layer was chosen to be that typically used in commercial production lines for wafer-to-wafer bonding. Figure 1b shows the eigenmode of this system, calculated using the CST software via our model depicted in Fig. 1a, while Fig. 1c denotes the confinement factors dependence on the $SiO_2$ thickness calculated via the supermode theory using MATLAB.

From these figures, we see that even for a large silicon waveguide width almost all of the mode is in the III-V active region. Therefore, this III-V stack design is not suitable for use in the supermode hybrid laser design described above and requires modifications. The ideal field distribution is one in which the mode resides as fully as possible in the III-V waveguide within the gain section of the laser and is then transferred as efficiently as possible into the silicon waveguide. In this way, the gain is effectively maximized when the mode resides mostly in the III-V stack, while the losses are minimized when the mode is transferred to the Si waveguide by increasing its width. Figure 1c further shows that simply reducing the thickness of the $SiO_2$ layer between the active region and the silicon waveguide does not significantly increase the proportion of the mode in the silicon waveguide. Therefore, the effective refractive index of the III-V stack must be reduced to enable the mode to be shifted down into the silicon waveguide when needed.

The effective refractive index of the III-V waveguide can be reduced by lowering the refractive indices of the confining and spacer layers within the active region, decreasing the number of quantum dot layers, or narrowing the width of the III-V waveguide. To simplify the fabrication and minimize cost, it is advisable to avoid varying the width of the III-V stack along its length [32, 33]. We, therefore, fix the width of the III-V stack to 4 μm (an effectively very large width compared to the silicon waveguide width) and we vary the number of quantum dot layers and change the materials used as barriers and cladding layers in the structure. More specifically, we select the appropriate ternary alloys to change the corresponding refractive indices of these layers so as to achieve the desired effective refractive index for the III-V waveguide.

Optimizing these parameters is challenging: they must be tuned to ensure that the effective refractive index of the III-V waveguide is low enough for the supermode to primarily reside in the silicon waveguide when the silicon width is large, yet still high enough to confine a significant portion of the mode within the III-V active region—rather than in the III-V cladding—when the silicon width is small. In this way, the gain is effectively maximized when the mode resides mostly in the III-V stack, while the losses are minimized when the mode is transferred to the Si waveguide by increasing its width. To achieve this fine tuning of the effective index while taking into account the above trade-offs, we proceed as follows.

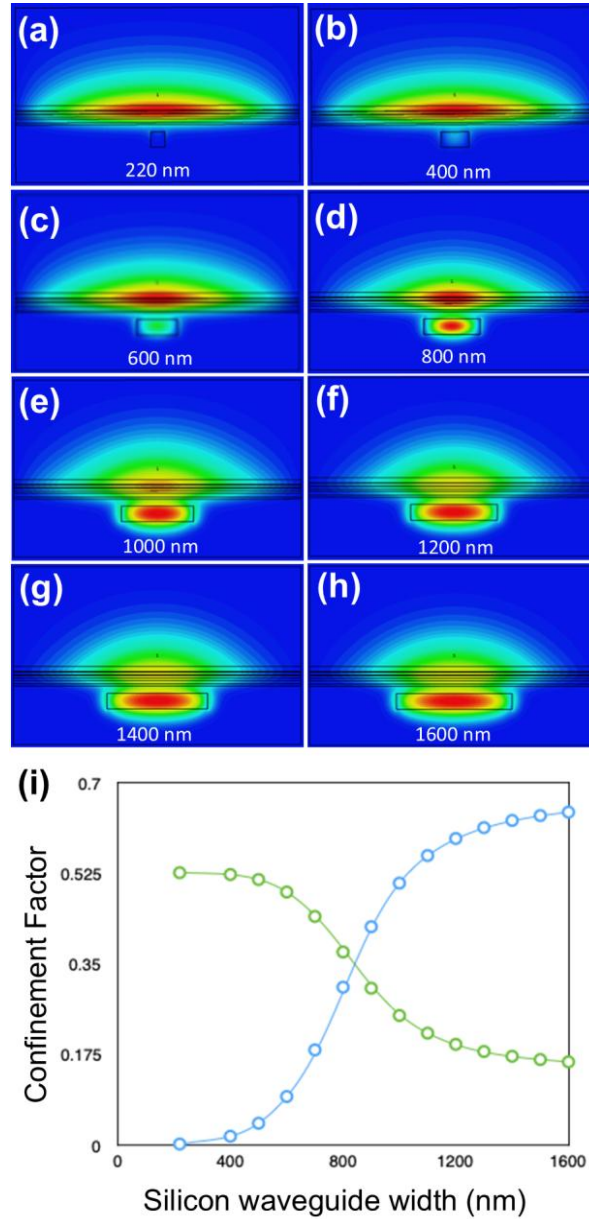

**Figure 2.** (a-h) 2D mode profiles obtained from CST simulations of the fundamental mode for silicon waveguide widths of 220-1600 nm. Each panel has its own arbitrary color scale. (i) Dependence of confinement factors of the silicon and III-V waveguides on the silicon waveguide width. Each point in (i) corresponds to a single deterministic simulation at the stated Si width (no averaging). Error bars are not shown because the solver is deterministic; uncertainty relevant to fabrication/material tolerances is quantified in Section 3.3 and Fig. 5.

For each specific design, once the cladding refractive index is selected, we systematically reduce the III-V stack's refractive index in a step-by-step manner by changing the composition of the barrier and waveguide layers in successive simulations (see Tables 1 and 2 for two typical structures). Changing the number of QD layers lowers the gain, and it is, therefore, left as the last resort in each design to lower them the least possible. There are, however, limits as to how much the composition of the barrier and waveguide layers can be adjusted. Growing high quality samples with an Al-content higher than x=0.45 is difficult with MBE, and such layers oxidize quickly in air. Having high-quality layers is crucial, especially within the active region of the stack. In addition, previous studies have shown that high-Al containing $Al_xGa_{(1-x)}As$ barriers shift the emission wavelength of the QDs away from the desired 1.3 μm [6]. Taking account of these considerations and our simulation results, we set the aluminum content limit for the barriers to x=0.375.

Having found that the refractive index of the III-V stack was still too high, we then started lowering the number of QD layers, until the desired mode profiles could be obtained. This was achieved for 2 QD layers in the final optimized epitaxial structure for this design, which is summarized in Table 2. In addition, we note that the silicon dioxide bonding layer thickness should also be as small as possible, to strengthen the waveguide coupling factor. We set it to 100 nm, a realistic value achieved in recent experimental investigations [2].

Figure 2(a–h) shows the 2D cross-sectional mode profiles of the fundamental mode, obtained using CST, for silicon waveguide widths ranging from 220 nm to 1600 nm. The silicon waveguide width (in nm) is shown in each plot. In each case, the full width of the structure is 4 μm and we envisage that the whole structure in Table 2 will be surrounded by silicon dioxide, or silicon nitride. The silicon dioxide passivates crystal dangling bonds at the surfaces avoiding surface current conduction bypassing the active region. Moreover, it crucially provides hermetic sealing to prevent the oxidation of AlAs and other high aluminum content semiconductor layers and to prevent unwanted ingress of exterior atoms diffusing into the active region. The other parameters of the system are given in Table 2. As can be seen from these graphs, most of the mode resides in the III-V waveguide for small silicon waveguide widths, less than 600 nm, while for large widths, more than 1000 nm, the mode shifts down into the silicon waveguide. Therefore, this design enables the realization of the desired mode profiles required for a supermode laser.

It should be noted that another way to achieve the desired mode profiles can be achieved without modifying the original III-V stack (Table 1), by increasing the thickness of the silicon waveguide above 220 nm, when the thickness of the silicon dioxide between the gain stack and the silicon waveguide is 100 nm or less. However, departing from 220 nm Si width would limit compatibility with CMOS technologies, and so we do not present these results in this paper.

With the parameters of the III-V stack chosen, we now turn our attention to the silicon waveguide design. The key parameters of the silicon waveguide are the width of the central waveguide, the width of the waveguide ends, and the length of the adiabatic taper. The structure, Figure 3, also includes a lateral Bragg grating at each end, compatible with 1.3 μm central wavelength section. The central Si waveguide, must be narrow enough for most of the mode to reside in the III-V central region, where gain occurs. However, it must also be kept wide enough for the fundamental mode of the silicon waveguide to have the same polarization as that in the III-V, which ensures that the desired supermode exists.

Both waveguides predominantly support the fundamental TE mode in the operating wavelength range, which ensures polarization matching and the existence of the desired supermode. The calculated mode overlap integral between the III-V and Si TE eigenmodes exceeds 90% in the central region, confirming that polarization alignment is well preserved.

Table 2. Final epitaxial structure proposed based on simulation and design trade-offs

|  | Material | Thickness | n | Repeats |
|---|---|---|---|---|
| p-type | $Al_{0.375}Ga_{0.625}As$ p+ 1e18 | 50 nm | 3.2 |  |
| Cladding | AlAs | 1400 nm | 2.91 | 1 |
| Active Region | $Al_{0.375}Ga_{0.625}As$ waveguide | 70 nm | 3.2 |  |
|  | $In_{0.18}Ga_{0.82}As$ capping | 2 nm | 3.434 |  |
|  | InAs | 2.7 ML | 3.532 | 2 |
|  | $In_{0.18}Ga_{0.82}As$ capping | 6 nm | 3.434 |  |
|  | $Al_{0.375}Ga_{0.625}As$ barrier | 37.5 nm | 3.2 |  |
|  | $Al_{0.375}Ga_{0.625}As$ waveguide | 32.5 nm | 3.2 |  |
| n-type | $Al_{0.375}Ga_{0.625}As$ n+ 5e18 | 50 nm | 3.2 |  |
| Bonding Layer | $Al_2O_3$ | 30 nm | 1.75 |  |
|  | $SiO_2$ | 100 nm | 1.45 | 1 |
|  | Si Waveguide | 220 nm | 3.5 |  |
| SOI | $SiO_2$ | 500 nm | 1.45 |  |
|  | Si Substrate | 300 μm | 3.5 |  |

*3.2. Silicon waveguide design*

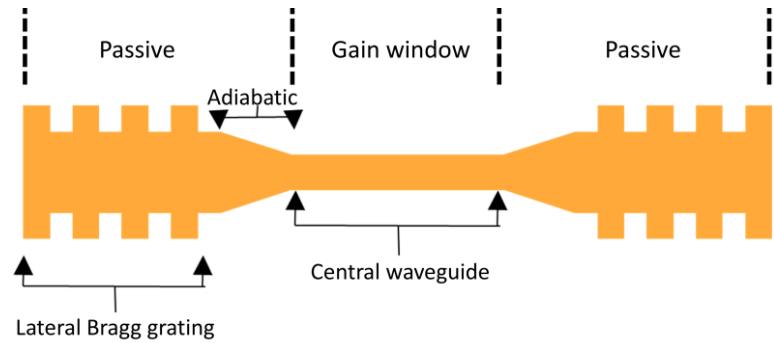

**Figure 3.** Schematic top view of silicon waveguide. The current injection region in the III-V waveguide (top) spans the Si central waveguide region, while the adiabatic tapers and lateral Bragg gratings are passive sections (no contacts; no current flows in the III–V). The central waveguide is 600 μm long; each adiabatic taper is 400 μm long, and each Bragg grating is 200 μm long.

The shape of this waveguide can then be changed adiabatically by increasing the width of the silicon waveguide to shift the mode into it in the desired regions. The width of the waveguide at the ends must be large enough for most of the mode to reside in the Si. It is also desirable to keep the difference between the central width and the width of the ends as small as possible to limit the length of the adiabatic taper, reducing losses in the system. Based on the confinement factors that we calculated, shown in Figure 2(i), and the requirement on the polarization of the fundamental silicon waveguide eigenmode, we set the central silicon waveguide width to 300 nm and the end waveguide width to 1500 nm.

In the above, we have focused on the fundamental mode of the system. It should be noted that while higher order modes will also propagate in the system, each with different propagation constants, the above optimization of the adiabatic taper and Bragg gratings for the fundamental mode will allow it to be preferentially selected for amplification and, thus, dominate the gain competition. More specifically, to keep the hybrid supermode well defined, we operate the central Si waveguide in the guaranteed single-mode regime of our 220-nm thick SOI stack. Using an eigenmode calculation, we obtain a $TE_1$ cutoff width of $w_c$ = 460 nm for our structure at $\lambda_0$ = 1.3 µm on 220-nm thick SOI, so we choose a Si waveguide width, $w_c$ = 300 nm to operate well within this window. This ensures that the central Si waveguide width in our design lies well within the guaranteed single-mode regime, supporting only the propagating mode $TE_0$. The end/DBR sections of the silicon waveguide are widened to $w_e$ = 1500 nm for efficient power transfer to Si; although the passive end grating sections are locally multimode, single-mode lasing is maintained by (i) the 400 µm adiabatic taper that is sufficiently long to ensure low inter-modal coupling ($|C_{mn}(z)| \ll 1$, Eq. (4)) and (ii) a weak, narrow-band lateral DBR that provides stronger feedback for the fundamental supermode than for higher order modes (HOM), which have different propagation constants (Fig. 6(b)) resulting in weaker feedback at the DBR. Figure 6(b) shows the effective refractive index as a function of silicon waveguide width for the three lowest-order modes of the Table 2 design. At a width of 1500 nm, the effective indices are $n_{eff}$ =2.995 ($TE_0$), 2.932 ($TE_1$), and 2.927 ($TE_2$). These simulations inform the design of the evanescent DBR mirrors so that $TE_0$ experiences stronger feedback, thereby dominating the modal gain competition and improving the side-mode suppression ratio (SMSR) of higher order modes [33].

For any HOM with detuning $\delta_m \equiv \beta_m - \beta_0$, we get

$$R_m(\delta_m) \propto \left(\frac{\kappa_{B,m}}{\delta_m}\right)^2 \qquad (3)$$

because the off-Bragg reflectivity $R_m(\delta_m)$ follows the $\left(\frac{\kappa_{B,m}}{\delta_m}\right)^2$ envelope obtained from Eq. (22) of Ref. 63 for $|\delta_m| \cdot L_g \gg 1$ and $|\delta_m| > \kappa_{B,m}$ ($\kappa_{B,m}$: mode-dependent Bragg coupling; $L_g$: grating length) [63]. Thus, only the fundamental supermode experiences net round-trip gain, ensuring that, although the end waveguide section of the hybrid laser is multimode, higher-order modes do not lase due to modal gain/loss discrimination (larger $g_{th,m}$ for higher order modes in Eq. (5)).

$$C_{mn}(z) \propto (dw/dz)\langle E_m | \partial \varepsilon / \partial w | E_n \rangle / (\beta_m - \beta_n)^2, \quad |C_{mn}(z)| \ll 1. \quad (4)$$

In Eq. (4), $C_{mn}(z)$ is the dimensionless inter-modal coupling coefficient from mode $n$ to $m$ due to the longitudinal change in geometry or materials, $E_m(x,y;z)$ is the transverse electric field of mode $m$ at longitudinal position $z$ (power-normalized), $\varepsilon(x,y;z)$ is the local relative permittivity, $w(z)$ the local Si core waveguide width, and $dw/dz$ is the taper slope [64]. The modal threshold gain for transverse order $m$ is given in Eq. (5).

$$g_{th,m} = \alpha_{i,m}/\Gamma_{active,m} + \left(1/(2 \cdot L_{cav} \cdot \Gamma_{active,m})\right) \cdot ln\left(1/(R_{1,m}R_{2,m})\right), (5)$$

where $\alpha_{i,m}$ is the internal loss, $\Gamma_{active,m}$ the active-region confinement factor, $L_{cav}$ the cavity length, and $R_{1,m}$, $R_{2,m}$ the end reflectivities [33].

With the Si waveguide geometry established, we now specify how these Si waveguide sections relate to the III–V ridge waveguide and to the current-injection footprint. The III–V waveguide is a uniform 4-µm-wide ridge that spans the central (gain) window, the adiabatic tapers, and the two laterally coupled Bragg reflectors (DBRs). However, only the central window is electrically pumped while the taper and DBR sections remain passive. In a practical implementation, this can be achieved by confining the metal overlap and vias to the central section and placing short isolation features (etched "moats" or implant-based current blocking) immediately upstream of each taper to prevent leakage of bias current into the passive ends. Feedback and mode transformation are implemented entirely in silicon: the end mirrors are silicon lateral Bragg gratings; the III–V above them remains unbiased and provides no gain. Thus, the cavity is a DBR cavity with passive Si mirrors and a localized hybrid III–V/Si gain section (Fig. 3). In these passive regions, the Si core is widened, thereby minimizing mode overlap with the unbiased III–V so keeping propagation losses low.

After sidewall smoothing, representative 220 nm thick SOI waveguides exhibit ≈ 1.5 dB/cm loss, so the passive cavity sections do not dominate the round-trip attenuation [2]. We note that the ≈ 1.5 dB/cm refers to a Si-only passive loss, with ≈ 100% confinement in the Si. In our structure, we can estimate this confinement to be ≈ 82% in Si and ≈ 18% residual in the unbiased III–V if we do not taper the III–V waveguide. In this case, we can estimate the additional loss, $\alpha_{pass}$ coming from the ≈ 18% residual overlap with the unbiased III–V as a weighted average of the Si and III-V waveguide losses to provide a maximum bound. Assuming a III-V loss of $\alpha_{III-V} \leq 10\ dB/cm$, and $\alpha_{Si} = 1.5\ dB/cm$, then $\alpha_{pass} = 0.82\ \alpha_{Si} + 0.18\ \alpha_{III-V} \leq 3.03\ dB/cm$. For the typical lengths of our design, with a 600 μm central region, 400 μm in each taper and 200 μm in each DBR, i.e., a total 1.2 mm passive section roundtrip, we estimate < 0.364 dB of extra attenuation. Assuming a $\alpha_{III-V} = 3\ dB/cm$; then $\alpha_{pass} = 1.77\ dB/cm$, the extra attenuation per round trip is 0.213 dB.

It is also interesting to clarify the mode evolution and modal content in both lateral directions, $x$ and $y$ ($w_c$ = 300 nm; $t_y$ = 220 nm; see Fig. 1(a) directions definition), in each waveguide section (Table 3), as well as the propagation and mode-expansion characteristics across the adiabatic transformers (Figure 4). Our analysis shows that the central gain window is multi-mode in $x$ and in $y$, with most modes being radiation unbound modes. In the $x$ direction, the first four modes are confined bound modes in the III-V waveguide, with the first three, $TE_0$, $TE_1$, and $TE_2$ shown in Figure 7. For robust laser operation, the tapers must transfer the fundamental optical mode from the III–V waveguide into the silicon waveguide with high efficiency. We, therefore, employed the principle of adiabatic coupling. An illustration of this strategy is depicted in Figure 2 and discussed in Section 3. This evolution is further illustrated in Fig. 6(b), which plots the effective indices of the three lowest-order modes versus Si width. Efficient coupling corresponds to a smooth evolution change in effective index from one value to another from w = 300 nm to w = 1500 nm, which is observed in Fig. 6(b). In contrast, $TE_1$ couples less efficiently as its effective index remains almost constant between w = 300 nm and w = 1500 nm which is to be expected since it is an antisymmetric mode.

Moreover, eigen-Mode Expansion (EME) simulations of the complete taper–gain–taper path show that the adiabatic tapers achieve fundamental-to-fundamental transfer with ≈ 98.5% $TE_0 \rightarrow TE_0$ coupling per transformer, i.e., from the hybrid $TE_0$ of the narrow, injected gain window to the Si-dominated $TE_0$ of the wide passive DBR section, while suppressing mode conversion coupling to higher-order lateral modes. Specifically, the integrated power in guided higher-order modes (HOMs) is ≈ 1.5%, and residual radiation is ≤ 0.5% as shown in Figure 4(b). Further modelling simulations will be required, in future work, to establish how much of the $y$-higher order bound mode power in the III–V waveguide is coupled into the fundamental mode in $y$ in the wider silicon waveguide end sections after the adiabatic tapered couplers. Any power that remains in III–V-localized $y$-higher-order modes traverses unbiased III–V in the passive sections and, therefore, incurs high propagation loss; moreover, these modes are not phase-matched to the Si narrowband DBR mirrors, so they cannot form a low-threshold cavity.

**Table 3.** Mode evolution and modal content in each waveguide section.

| Section | Si width | Lateral modes at 1.3 µm | Mode mostly in | Notes |
|---|---|---|---|---|
| Central gain window | $w_c$ = 300 nm | Multi-mode in $x$ and $y$ | III–V | Current injection |
| Adiabatic taper (each) | $w_c$ = 300 nm → 1500 nm (smooth) | Multi-mode in $x$: Only $TE_0$ and $TE_2$ supported[1] | Shifts from III–V to Si | $TE_0 \rightarrow TE_0$ > 98.5%; passive |
| DBR mirror section | $w_e$ = 1500 nm | Multi-mode in $x$: Only $TE_0$ and $TE_2$ guided | Predominantly (82%) in Si | Passive – no gain |

[1]Multi-mode in $x$ (even parity): only $TE_0$ and $TE_2$ are supported; odd modes ($TE_1$ and $TE_3$) are symmetry-forbidden from coupling.

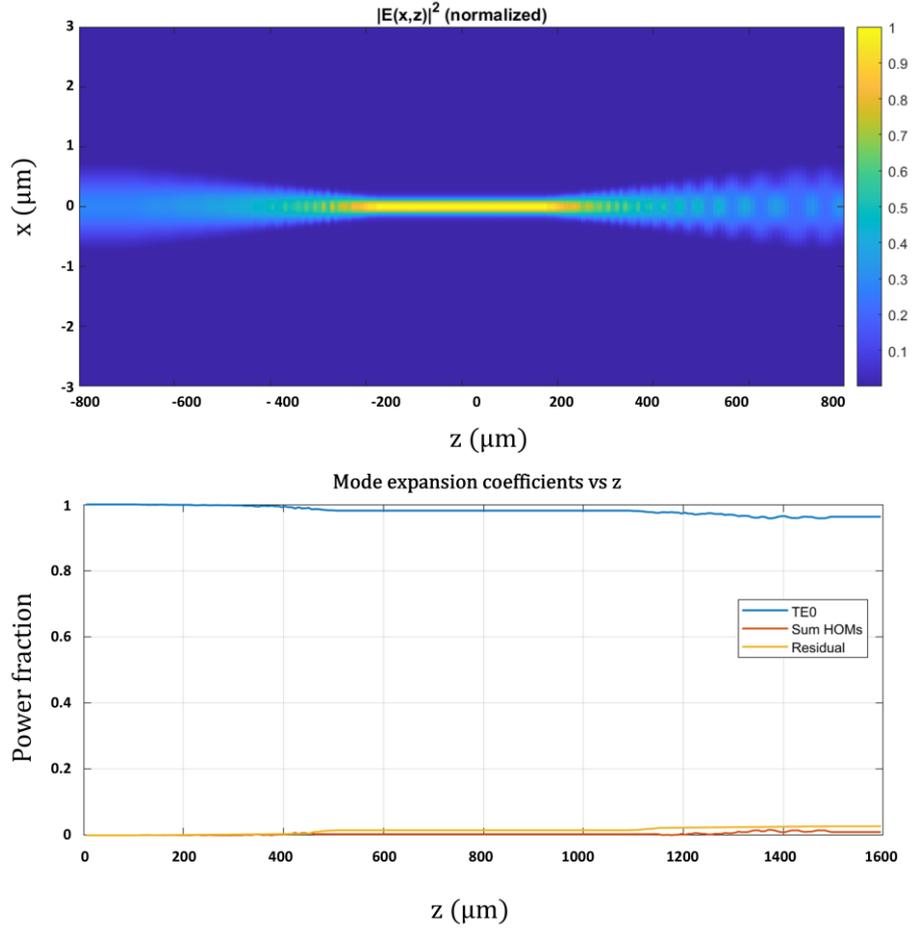

**Figure 4.** Propagation and mode-expansion across the adiabatic taper transformers. (a) EME-computed field-intensity map $|E(x,z)|^2$ ($x$–$z$ cross-section 300 nm x 220 nm) for a taper–gain–taper path. ($w_c$ = 300 nm; $w_e$ = 1500 nm; $L_{taper}$ = 400 µm; $z$ = 0 is the center of the gain section). The map shows the adiabatic transfer from the III–V-weighted hybrid $TE_0$ in the central gain window to the Si-weighted hybrid $TE_0$ in the passive sections. (b) Running modal power fractions vs. propagation distance $z$: $TE_0$ (blue), sum of guided HOMs $TE_{m \geq 1}$ (red), and radiation (orange). The $TE_0 \rightarrow TE_0$ coupling per transformer is ≥ 98.5%, with HOMs ≤ 1.5% and radiation ≤ 0.5%.

3.3. Tolerances to growth and fabrication inaccuracies.

In practice, device fabrication inevitably involves some degree of layer inhomogeneity or roughness, meaning that the desired design cannot be precisely replicated to the intended dimensions and compositions due to imperfections introduced during the growth and fabrication processes [53, 54, 56]. Therefore, it is important to investigate how the obtained mode distribution varies if we change the key parameters slightly from their design values. Having this in mind, the width of the silicon waveguide has been chosen such that if the width of the waveguide in the central or end section differs by even as much as 100 nm from its design value, the confinement factors will change by less than 0.05, as seen in Fig. 2(i). Figure 5(a-h) shows how the confinement factors of the key regions in the structure depend on the design parameters. These dependencies were investigated for a silicon waveguide width of 300 nm Figure 5 (panels a, c, e, g) corresponding to the central region, and for a width of 1500 nm Figure 5 (panels b, d, f, h) corresponding to the end sections of the DBR laser. We note that because the results in Fig. 2(i) are deterministic single-point calculations, we do not plot statistical error bars; instead, Fig. 5 provides the relevant uncertainty envelopes via parameter-sensitivity curves.

By comparing Fig. 5a and b, we see that the confinement factors depend strongly on the thickness of the silicon waveguide for the 1500 nm wide waveguide. However, this should not be a problem since the CMOS foundries which produce the silicon layer from which the waveguide will be etched can achieve very fine control over its thickness, on the order of a few nm for a 220 nm waveguide [2]. Additionally, we see in Fig. 5c and d that the confinement factors depend strongly on the refractive index of the doped layers near the active region. This too should not be a problem because the material composition element ratios of these layers can be adjusted until the desired refractive index of 3.2 is obtained, thanks to $Al_xGa_{(1-x)}As$ being lattice-matched over its entire composition range [55].

Furthermore, we see a strong dependence of the confinement factors on the refractive index of the cladding, with the mode shifting upwards into the cladding when its refractive index is increased (Fig. 5e, f). For this reason, the design demonstrated in Table 2 (Design 2) employs an AlAs cladding (x = 1), which provides the lowest refractive index (n = 2.91) among $Al_xGa_{(1-x)}As$ alloys. However, we note that designs up to $Al_{0.7}Ga_{0.3}As$ with

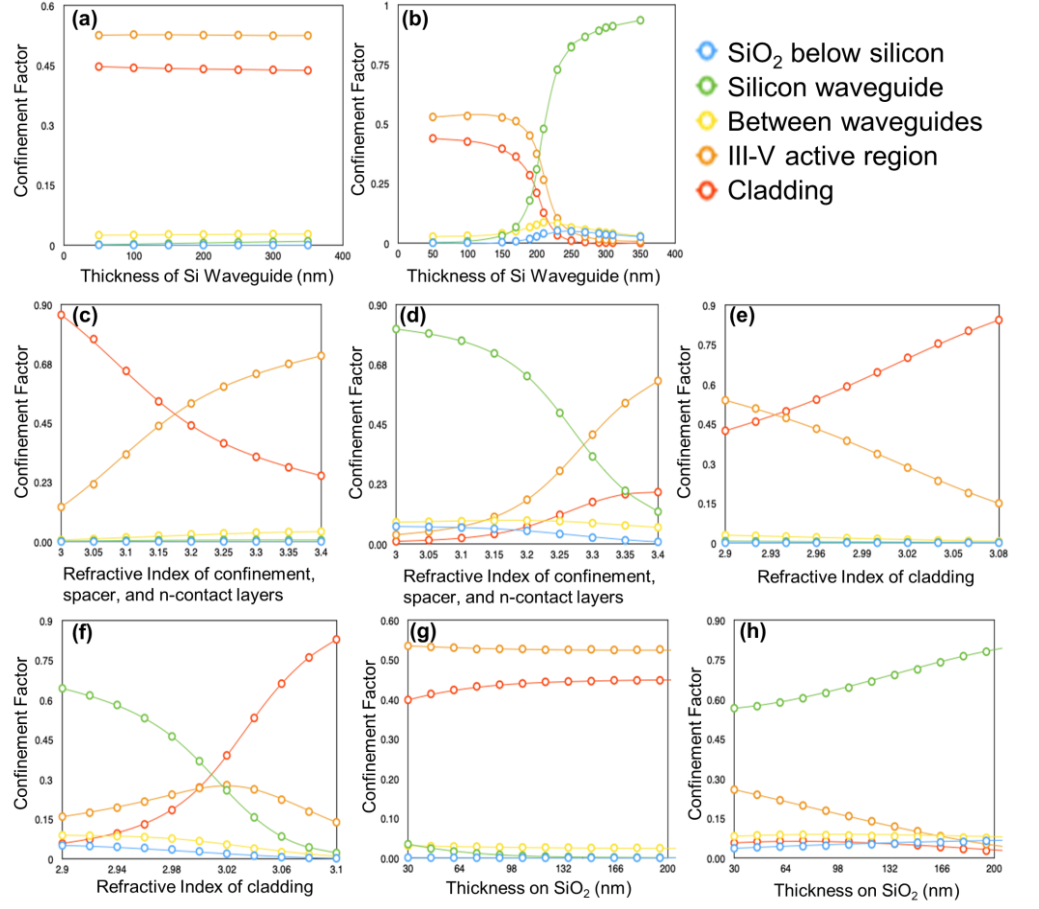

**Figure 5.** (a-h) Dependence of confinement factors for each region in the structure on design parameters listed in Table 2. The specific regions associated with each confinement factor are given by the colors of the circles, as specified by the top-right hand legend. For graphs (a, c, e, g) the silicon waveguide is 300 nm wide, whereas in panels (b, d, f, h) the silicon waveguide is 1500 nm wide.

n =3.02 remain viable (see Fig. 5e, f), offering some flexibility depending on growth capabilities and relative material quality. The trade-off between the $Al_{0.7}Ga_{0.3}As$ cladding design (Design 3) and Design 2 involves approximately 35% additional loss due to mode leakage into the $Al_{0.7}Ga_{0.3}As$ cladding in Design 3, versus potential losses from reduced material quality in AlAs in Design 2. The latter should be evaluated experimentally following the growth of both materials. Another interesting possibility is to also make the III-V waveguide layer slightly thicker in Design 3, helping to reduce leakage at the expense of reduced III-V/Si coupling.

Finally, in Fig. 5g, h, we see that the confinement factors do not depend strongly on the thickness of the silicon dioxide layer between the waveguides. However, increasing the distance between the waveguides will decrease the strength of the coupling between them, making processes that decrease the $SiO_2$ thickness preferable [2].

*3.4. Environmental sensitivity and open-system considerations.*

Beyond the closed-cavity analysis above, we consider the environmental aspects most relevant to hybrid III–V-on-Si quantum-dot lasers. The ambient temperature affects the Bragg condition primarily through silicon's thermo-optic response (its refractive index change dominates over its thermal expansion), which for our geometry leads to resonance/Bragg drifts of the order of 0.1 *nm K$^{-1}$*; to the first order:

$$\frac{d\lambda_B}{dT} \approx \lambda \cdot \left[\frac{\frac{\partial n_{eff}}{\partial T}}{n_{eff}} + \alpha_{Si}\right]. \quad (5)$$

Here, $\lambda_B$ is the Bragg wavelength, $n_{eff}$ the effective index of the guided supermode, $\frac{\partial n_{eff}}{\partial T}$ its thermo-optic coefficient, and $\alpha_{Si}$ the silicon linear thermal-expansion coefficient. In practice, we can remove slow temperature drifts with a compact thermal trim (e.g., a thermoelectric cooler, TEC) without altering the optical design. Parasitic back-reflections can form an external cavity; however, within standard semiconductor-laser feedback models, sufficiently weak returns preserve a unique, stable single-mode solution, whereas stronger or longer-delayed feedback can induce multi-stability and coherence collapse. Standard packaging therefore, targets high return loss (≈ 40–60 dB), for example using angled and/or AR-coated interfaces, and it avoids short external cavities so operation remains in the stable regime.

Coupling of the quantum-dot gain medium to finite-memory solid-state reservoirs (such as acoustic phonons and charge noise) can introduce non-Markovian dephasing and modify optical noise and linewidth; recent open-system studies in related platforms (giant-atom waveguide quantum electrodynamics and non-Markovian or universal photon blockade) clarify these mechanisms. While our device operates firmly in the many-photon laser regime—where such couplings primarily manifest as excess phase/amplitude noise rather than few-photon blockade—these phenomena are particularly relevant when analyzing single-photon sources based on similar designs [65-67]. Overall, the design choices developed above make the single-mode solution robust against these environmental perturbations.

4. Discussion

In this paper, we have demonstrated how the mode profiles in hybrid silicon lasers depend on the dimensions of the silicon waveguide as well as the dimensions and composition of the III-V gain stack, using a combination of 2D (MATLAB) and 3D (CST) simulations. Guided by knowledge acquired through these simulations, we developed a design strategy in which the III-V gain stack was systematically modified to achieve the desired mode profiles in a hybrid laser with a CMOS-compatible Si waveguide thickness of 220 nm. Importantly, the resulting mode profiles were found to be tolerant to deviations from design parameters that might be expected during fabrication.

The design presented herein (Table 2) requires a high-Al-content cladding layer, which is known to oxidize rapidly in air and therefore must be properly encapsulated following fabrication. Effective encapsulation can be achieved, for example, using silicon dioxide ($SiO_2$) or other dielectric coatings. In addition, we assessed the optical impact of this high-Al content cladding, which additionally lies in the p-doped region of our stack. While in some designs the doping of this layer near the active region can be a concern due to free-carrier absorption losses, our confinement factor calculations show that in our design this is not significant because of the limited overlap of the optical mode with the cladding. In practice, the effect of this layer can be further reduced by using a graded doping profile, such that the lower part of the cladding—where the mode overlap is non-negligible—remains more lightly doped, while the higher doping needed for electrical contact is applied only in the uppermost region. Our simulations can directly inform such an approach by identifying the depth at which the optical field becomes negligible.

We note that, should the growth or fabrication steps associated with this high-Al-content cladding prove challenging to control, alternative designs can be explored by making only slight modifications to our final proposed structure. One such design is discussed in Section VI. Another appealing option is to consider width tapering the III-V waveguide, which increases flexibility in the choice of compatible cladding materials. So far, we have focused on fixed-width gain stacks, as we believe this simplifies fabrication. However, while tapering introduces fabrication challenges, it offers an alternative means of lowering the effective refractive index of the gain region, potentially reducing the need to modify its composition or the number of QD layers to achieve the desired mode profiles.

With this in mind, one could envision a design using the original III-V stack described in Table 1, combined with a 220 nm-thick silicon waveguide of fixed width, where tapering of the III-V stack is used to push the optical mode into and out of the silicon. The benefit of introducing a taper to the III-V waveguide has been demonstrated by Morais et. al. while using a 300 nm thick Si waveguide [28]. Even though the authors reported occasional bubbles resulting from the transfer printing process, the lasers exhibited a relatively low threshold current of 47 mA at room temperature and 78 mA at 80°C [28]. Although a detailed analysis of the III-V taper is beyond the scope of this paper, we performed preliminary simulations to quantify its potential benefits in mode engineering. Specifically, we simulated how the confinement factors are affected by altering the width of the III-V waveguide for the design of Table 2. The results are shown in Figure 6(a) alongside the simulations for the Si waveguide dependence for comparison. Notably, our results indicate that reducing the width of the III-V waveguide to 500 nm will leave less than 4% of the mode in the active region.

Our results also suggest that with an effective III-V taper the design of Table 1 also becomes viable, and we plan to investigate this further in future work. Finally, another possible solution is to remove the cladding layer altogether; however, this would result in a greater portion of the optical mode residing in the highly doped and metal contact layers, leading to significantly higher optical losses. In such a case, it may be worthwhile to explore combining it with a lateral injection scheme.

Our results highlight that InAs/GaAs quantum dots can be used with 220-nm thick SOI waveguides while maintaining favorable hybrid supermode confinement and single-mode selection. Quantum dots are particularly attractive for silicon photonics because of their temperature-insensitive operation and low threshold currents, as recently underscored in experimental Si-based laser demonstrations [6, 28]. In a complementary direction, QD–2D-material heterostructures have achieved high-performance LEDs [68], highlighting the broader relevance of QD emitters for integrated photonics and motivating integration-oriented future studies.

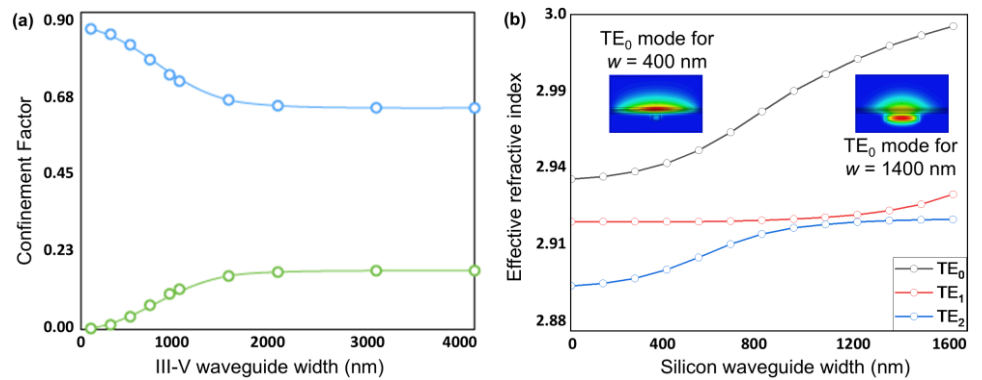

Figure 6. (a) Confinement factor dependence on III-V waveguide width for the design of Table 2, with the Si waveguide width fixed at 1500 nm (end sections). Green and blue circles represent the III-V active region and Si waveguide respectively. (b) Effective refractive index versus silicon waveguide width for the three lowest-order modes of the Table 2 design— $TE_0$, $TE_1$, and $TE_2$ (black, red, and blue, respectively).

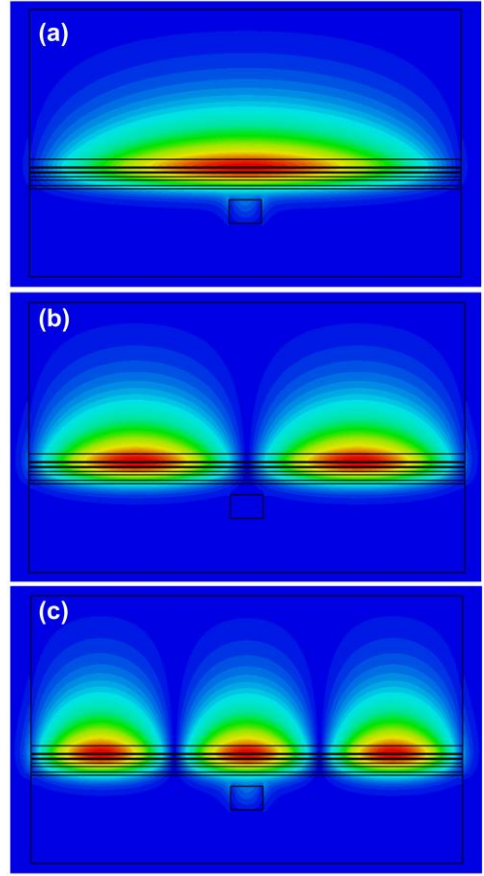

Figure 7. 2D mode profiles, in which each lobe has a π radians phase shift relative to the adjacent lobes, obtained from CST simulations of the first three confined modes for silicon waveguide width 300 nm. (a) fundamental mode, $TE_0$; (b) $TE_1$; (c) $TE_2$

A potential practical realization of the design presented herein, follows the now-standard heterogeneous silicon photonics workflow [2]. Silicon waveguides and the lateral Bragg gratings are first patterned on SOI using standard CMOS lithography; wafer-bonding of an unpatterned III–V epitaxial gain film onto the processed SOI is then performed (direct plasma-activated or adhesive), after which the III–V substrate is removed and the III–V ridge, contacts, and isolation are defined [2]. This route enables single-frequency hybrid lasers in which the Bragg mirrors are etched in silicon (DBR/DFB) with the gain only in the III–V material, matching our cavity mode partitioning. The III-V layer can also be shaped lithographically being wider in the central region and tapering down and vanishing in the grating regions.

Laterally coupled gratings have likewise been demonstrated in both III-V and hybrid III–V/Si lasers, confirming exceptionally high side mode suppression ratio (SMSR) [32]. Adiabatic tapers (single or double) between the III–V and Si guides are routinely used to achieve efficient and alignment-tolerant mode transfer over tens of micrometers, consistent with our coupling strategy [2, 3]. Finally, quantum-dot gain has been heterogeneously integrated on SOI in electrically pumped lasers, including widely tunable O-band devices with high SMSR and narrow linewidth [15, 16, 18, 28], underscoring that our materials stack, cavity partitioning, and coupling approach are compatible with present-day experimental directions.

## 5. Conclusions

In this work, we demonstrated a method for applying supermode theory to practical hybrid laser structures and used it to investigate the key trade-offs involved in their design. Our analysis focused particularly on the challenges of integrating quantum dot-based hybrid lasers with CMOS-compatible 220 nm-thick silicon waveguides—an important step toward temperature insensitive Si-based lasers for scalable silicon photonics.

Based on this analysis, we proposed a novel structure that addresses the coupling challenges between III-V quantum dot active regions and thin Si waveguides. We also identified potential fabrication challenges and proposed alternative designs that require only minor modifications to the final structure, while relying on the same theoretical framework and trade-off considerations.

We believe these alternative approaches merit further theoretical and experimental exploration, and we hope that the modeling strategy and design guidelines presented here will serve as a solid foundation for future research in hybrid laser integration.

**Author Contributions:** Conceptualization, D.S and K.P.; methodology, D.S and K.P.; software, R.S.; validation, R.S., D.S and K.P.; formal analysis, R.S.; investigation, R.S. and K.P; resources, D.S.; data curation, R.S.; writing—original draft preparation, R.S. and K.P; writing—review and editing, D.S and K.P.; visualization, R.S.; supervision, D.S.; project administration, D.S. and K.P; funding acquisition, D.S. All authors have read and agreed to the published version of the manuscript.

**Funding:** This research was funded by L3MATRIX. The L3MATRIX project is co-funded by the Horizon 2020 Framework Programme of the European Union with Grant Agreement Nr. 688544. L3MATRIX project is an initiative of the Photonics Public Private Partnership. P. R. Smith thanks the EPSRC for funding via the UCL – Cambridge Centre for Doctoral Training in Integrated Photonic and Electronic Systems.

**Data Availability Statement:** All data are available upon request.

**Acknowledgments:** The authors thank Professor Huiyun Liu, Professor Alwyn J. Seeds, Dr James Seddon, and Dr Michele Natrella of the Photonics Research Group for helpful discussions.

**Conflicts of Interest:** The authors declare no conflicts of interest.